\documentclass{aastex}
\usepackage{emulateapj5}
\usepackage{epsf}
\usepackage{psfig}
\usepackage{graphicx}
\usepackage{lscape}
\usepackage{natbib}
\usepackage{enumerate}

\usepackage{rotating}

\begin{document}

\title{Characterizing the Rigidly Rotating Magnetosphere Stars HD 345439 and HD 23478$^{10}$}
\author{J. P. Wisniewski\altaffilmark{1}, S. D. Chojnowski\altaffilmark{2}, J. R.A. Davenport\altaffilmark{3}, J. Bartz\altaffilmark{4}, J. Pepper\altaffilmark{4}, 
D. G. Whelan\altaffilmark{5}, S. S. Eikenberry\altaffilmark{6}, J. R. Lomax\altaffilmark{1}, S. R. Majewski\altaffilmark{7}, N.D. Richardson\altaffilmark{8,9}, 
M. Skrutskie\altaffilmark{7}}

\altaffiltext{1}{{Homer~L. Dodge Department of Physics \& Astronomy, 
                 The University of Oklahoma, 440 W. Brooks Street, 
                 Norman, OK 73019, USA};
                 {\tt wisniewski@ou.edu}}
\altaffiltext{2}{Department of Astronomy, New Mexico State University, 1780 E University Ave, Las Cruces, NM 88003}
\altaffiltext{3}{Department of Astronomy, University of Washington, Box 351580, U.W., Seattle, WA 98195}
\altaffiltext{4}{Lehigh University, Department of Physics, 413 Deming Lewis Lab, 16 Memorial Drive, East Bethlehem, PA 18015}
\altaffiltext{5}{Department of Physics, Austin College, 900 N. Grand Ave., Sherman, TX 75090, USA}
\altaffiltext{6}{Department of Astronomy, University of Florida, 211 Bryant Space Science Center, Gainesville, FL 32611}
\altaffiltext{7}{Department of Astronomy, University of Virginia, P.O. Box 400325, Charlottesville, VA 22904-4325, USA}
\altaffiltext{8}{D\'epartement de physique and Centre de R affil echerche en Astrophysique du Qu\'ebec (CRAQ)}
\altaffiltext{9}{Universit\'e de Montr\'eal, C.P. 6128, Succ.~Centre-Ville, Montr\'eal, Qu\'ebec, H3C 3J7, Canada}
\footnotetext[10]{This publication is partially based on observations obtained with the Apache Point Observatory 3.5 m telescope, which is owned and operated by the Astrophysical Research Consortium.}


\begin{abstract} 
The SDSS III APOGEE survey recently identified two new $\sigma$ Ori E type candidates, HD 345439 and HD 23478, 
which are a rare subset of rapidly rotating massive stars whose large (kGauss) magnetic fields confine circumstellar material 
around these systems.  Our analysis of multi-epoch photometric observations of HD 345439 from the KELT, SuperWASP, and ASAS 
surveys reveals the presence of a $\sim$0.7701 day period in each dataset, suggesting the system is amongst the faster known 
$\sigma$ Ori E analogs.  We also see clear evidence that the strength of H$\alpha$, H I Brackett series lines, and He I lines also 
vary on a $\sim$0.7701 day period from our analysis of multi-epoch, multi-wavelength spectroscopic monitoring of the system from 
the APO 3.5m telescope.  We trace the evolution of select emission line profiles in the system, and observe coherent line profile 
variability in both optical and infrared H I lines, as expected for rigidly rotating magnetosphere stars.  We also analyze the 
evolution of the H I Br-11 line strength and line profile in multi-epoch observations of HD 23478 from the SDSS-III APOGEE instrument.  
The observed periodic behavior is consistent with that recently reported by Sikora and collaborators in optical spectra.  

\end{abstract}


\section{Introduction}
\label{sec:intro}

Although they lack a convective outer envelope, a subset of massive rapidly rotating stars, aka $\sigma$ Ori E analogs, exhibit strong magnetic fields 
(10s of $kG$ \citealt{lan78,oks10,riv10,gru12}), which act to confine their winds into a circumstellar disk.  {The Rigidly Rotating Magnetosphere (RRM) model \citep{to05b,tow05,tow07,ok15b} is the leading theoretical framework to explain the distribution of matter in these systems.  While 
the RRM model, along with accounting for the effects of an inhomogeneous distribution of elemental abundances in the photosphere, does explain many of the qualitative features of these 
systems such as the general periodic behavior of line profile morphologies, light curve variations, and longitudinal magnetic field variations, discrepancies in the observed versus predicted line profile morphologies and light curves at specific phases suggest that model is missing important physics.  As suggested by \citet{ok15b}, details such as additional scattering that takes place around 
the magnetosphere and the shape of high density concentrations of disk material (see also \citealt{car13}) may need to be better described within models to achieve better agreement with 
observational data.  Attempts to further refine our understanding of $\sigma$ Ori E analogs are hampered by the relatively small numbers of such systems 
identified to-date, and the sparse availability of multi-epoch datasets for confirmed analogs.  Moreover, the availability of robust multi-wavelength 
datasets for such systems remains limited, especially in the infrared where only a few exploratory investigations have been performed \citep{eik14,oks15}.

\citet{eik14} and \citet{cho15} recently identified at least two new likely $\sigma$ Ori E analogs via analysis of SDSS III/APOGEE \citep{eis11} spectroscopic observations, thereby increasing the number of such systems known by $\sim$10\%.  One of these candidate analogs, HD 345439, is a He rich B2V star that was reported to have $v \sin i$ = 270 $\pm$ 20 km s$^{-1}$, making it one of the fastest rotating RRM systems observed \citep{eik14}.  More recently, 
\citet{hub15} used VLT/FORS2 spectropolarimetric observations to explore the magnetic field properties of the system and while no 
magnetic field was found in the total dataset, the longitudinal magnetic field varied by 1$kG$ over the 88 minutes of integration, 
suggesting the presence of a rapidly varying magnetic field.  \citet{hub15} suggest this could be consistent with HD 345439 having a rotation period significantly shorter than 1 day.  The other candidate analog, HD 23478, is a He normal B3IV star with a reported 
$v \sin i$ = 125 $\pm$ 20 km s$^{-1}$, and a photometric period of 1.0498 days \citep{jer93,sik15}.  Recent spectropolarimetric 
observations of the system have confirmed the presence of a weakly variable longitudinal magnetic field of $<$ B$_{z}$ $>$ $\sim$ -2.0 $kG$ in the system, along with periodically variable H$\alpha$ emission and He I absorption \citep{sik15}, indicating the system is indeed a $\sigma$ Ori E analog.

In this paper, we present multi-epoch optical and infrared spectroscopic and optical photometric observations of HD 345439 and multi-epoch infrared spectroscopic observations of HD 23478.
These data help to identify likely rotational periods (HD 345439),  characterize the confinement of material in their circumstellar disks (HD 345439 and HD 23478), and detail the level of He I 
variability present (HD 345439) that could help diagnose surface abundance anomalies when combined with future detailed modeling.  

\section{Observations}
\label{sec:observ}
\subsection{Optical Spectroscopy}
Seventeen epochs of optical spectra of HD 345439 were obtained using the Astrophysical Research Consortium Echelle Spectrograph (ARCES) on the 
Apache Point Observatory's (APO) $3.5$m telescope between 2012 September 2 and 2014 June 13 (see Table 
\ref{tab:observation_table}).  ARCES \citep{wan03} is a high resolution, cross-dispersed visible light 
spectrograph \footnote[10]{http://www.apo.nmsu.edu/arc35m/Instruments/ARCES/} that obtains R$\sim$31,500 
spectra between 3600-10000 $\AA$.  We observed bias, flat field, and ThAr lamp exposures for the echelle on every night.  
These data were reduced using standard techniques in IRAF\footnote[11]{IRAF is distributed by the National Optical Astronomy Observatories, which are operated by the Association of Universities for Research in Astronomy, Inc., under cooperative agreement with the National Science Foundation.}, and individual spectral lines of interest were continuum normalized using the IRAF task 
\textit{continuum}.  

\subsection{IR Spectroscopy}
We obtained six epochs of IR spectra of HD 345439 from 0.95 - 2.46$\mu$m using Triplespec on APO's 3.5m telescope on 2013 
July 14 (see Table \ref{tab:observation_table}).  
These data were obtained with the 1$\farcs$1 x 43$\farcs$0 slit in standard ABBA observing mode, yielding R$\sim$3500 
spectra.  We obtained spectra of the nearby A0V star HD 189690 after each observational sequence of HD 345439 to facilitate
accurate telluric correction.  These data were reduced and telluric corrected using Triplespectool, a modified version of 
Spextool \citep{vac03,cus04}.

We also analyzed sixteen epochs of high resolution (R$=22,500$), H-band spectra (1.5145--1.6960 $\mu$m) of HD 23478 (see Table \ref{tab:observation_table}) observed with 
the 300-fiber APOGEE instrument on the SDSS 2.5m telescope\citep{gun06}, and released as part of the SDSS data release 12 (DR12) \citep{ala15}.    
The reduction of APOGEE data and details of the SDSS data release 12 (DR12) spectra used are described in \citet{nid15} and \citet{hol15}.  The data for HD 23478 from the ``blue'' and ``green'' APOGEE detectors \citep{nid15} exhibit clear signatures of saturation and were excluded from our analysis.

\subsection{Optical Photometry}
We analyzed 2392 photometric observations of HD 345439 obtained by the Kilodegree Extremely Little Telescope (KELT) North 
transit survey \citep{pep07,siv09} taken between 2007 May 30 and 2013 June 13. 
KELT consists of a 42 mm lens imaging a 26$^{\circ}$ $\times$ 26$^{\circ}$ field of view onto a 
4000 $\times$ 4000 pixel CCD. KELT uses a red-pass filter with a 50\% transmission point at 490 nm, which, when folded with the CCD response, yields an effective bandpass similar to R, but broader (see Section 2.1 of 
\citealt{pep07}).  After flat-fielding, relative photometry was extracted using a heavily modified version of 
the ISIS image subtraction package \citep{ala98}, combined with point-spread fitting photometry using DAOPHOT \citep{ste87} (see Section 2.2 of \citealt{siv09}). The final KELT light curve had a typical relative photometric precision of ~2\%.

We supplemented these KELT photometric observations of HD 345439 with archival photometric data from the Super Wide Angle 
Search for Planets (SuperWASP; \citealt{but10}) and the All Sky Automated Survey (ASAS; \citealt{poj97}).  SuperWASP observed 
the system over a span of 67 nights in 2007, providing 979 epochs of photometry in a broad white light filter, with an average 
photometric uncertainty of 0.02 magnitudes.  ASAS obtained 221 epochs of photometric observations of the system from 2003 - 2009 
in the I-band, with a typical average photometric uncertainty for HD 345439 of 0.04 magnitudes.

\section{Results}
\label{sec:results}
\subsection{Photometry: HD 345439}\label{Be50photdiscussion}

\citet{eik14} suggested that HD 345439 was among the most rapidly rotating $\sigma$ Ori E-type stars based on its 
observed $v \sin i$ value (270 $\pm$ 20 km s$^{-1}$).  We further explore the rotational rate of this star by examining all of the available 
KELT photometry for the system, and computing Lomb-Scargle (LS) periodograms \citep{lom76,sca82} of the data to search for 
evidence of periodicity.  We searched for periodic signatures both in the individual KELT seasons as well as in the full
KELT light curve in the period between 0.1 and 5.0 days.  The most prominent LS peak occurred 
at 0.7701 $\pm$0.0003 days (Figure \ref{kelt_fig}); we note that other suggested periods disappear after the data is whitened.

Phase-folding our KELT photometry by this suggested 0.7701 day period (Figure \ref{kelt_fig}) and binning these data, we see the clear 
signature of small amplitude ($\sim$22 mmag) periodic variations that rise above the errors of each bin (mean binned error $\sim$16 mmag).  The zero point of the ephemeris, with phase = 0 corresponding to the first minimum in the 
lightcurve, occured at JD = 2454251.3432.

We also analyzed the available ASAS and SuperWASP archival photometry of the system in a similar manner.  We found a period of 0.7695 $\pm$0.0078 days in the SuperWASP data (Figure \ref{kelt_fig}).  
Folding these data by this phase and computing median flux values in bins of 0.05 phase produced the phased-folded light curve for the SuperWASP data shown in Figure \ref{kelt_fig}.  Computing a LS periodogram for the ASAS data revealed evidence of a periodic 
signal at 0.7702 $\pm$ 0.0001 days (Figure \ref{kelt_fig}).  To aid the visual interpretation of this light curve we again computed median 
flux values in bins of 0.05 phase and overlayed these median values on the ASAS phase-folded light curve shown in 
Figure \ref{kelt_fig}.  Note that the mean binned error ranges from $\sim$16 mmag (SuperWASP) to $\sim$27 mmag (ASAS).  

Each individual photometric dataset for HD 345439 yields the same LS peak signal (to within $\sim$50 seconds).  Moreover, 
the phase-folded light curves of the SuperWASP, ASAS, and KELT data (Figure \ref{kelt_fig}) exhibit the same basic morphology, albeit with low SNR for the ASAS data.  This consistency indicates the complex morphology observed 
is in fact real, and the period of the system is $\sim$0.7701 days.

\subsection{Spectroscopy: HD 345439}\label{Be50specdiscussion}

We compute equivalent width (EW) line strengths for select major lines from continuum normalized data.  Because the observed line profiles were often complex, we computed EWs by intregrating 
over the full profile of each line rather than fitting each profile with a functional form.  We compile all of our measured 
optical and IR EWs in Table \ref{tab:observation_table}, along with their errors, computed using the technique discussed in 
\citet{cha83}.  Although not included in the formal error bars, we suggest that continuum normalization placement uncertainties could add an additional $\sim$5\% to our quoted errors.

The EWs of numerous emission and absorption lines in HD 345439 exhibit clear evidence of variability which is phase-dependent.  
Our optical spectra for HD 345439 cover 60\% of the phase-space of the system; the periodic behavior of the 
H$\alpha$ EW is indicative of circumstellar gas confined by the system's suspected magnetic field (Figure \ref{Be50phase}).  Although 
comparatively our IR data cover a smaller portion of phase-space, these data (e.g. Br-11, Figure \ref{Be50phase}) exhibit a 
similar general trend with phase as for H$\alpha$.  We find weak evidence for periodic behavior in the He I 4144, 4387, 4921, 5015, and 6678 \AA\ lines 
(a subset of these lines are shown in Figure \ref{Be50phase}), albeit with opposite amplitude as observed in H$\alpha$.  Whereas the observed H$\alpha$ EW variations 
likely arise from magnetospherically confined gas, determining the cause of the He I EW variations is more challenging.  When He I EW variations exhibit clearly different trends as a function 
of phase than H$\alpha$, this has been attributed to differences in photospheric surface abundances (see e.g. \citealt{gru12}).  However when He I EW variations exhibit similar trends as 
a function of phase as H$\alpha$, such behavior could arise both from photospheric surface abundance differences and due to occultations of the photosphere by the circumstellar 
disk \citep{tow08, oks12,riv13}.  Determining the role of anisotropic photospheric surface abundances and occultations by the disk in forming HD 345439's observed He I EW 
behavior will require additional observations that enable elemental abundance mapping (e.g. \citealt{ok15b}) and detailed modeling of the system. 

We further explore HD 345439's circumstellar material by investigating dynamical spectra of our data.  We first consider our data from 2013 July 14, 
whereby we obtained near-contemporaneous optical and IR time series spectra of HD 345439 throughout $\sim$40\% of its orbital phase.  Figure 
\ref{Be50halpha} illustrates the dynamical spectrum for the H$\alpha$, H I Br-11, and H I Br-$\gamma$ lines; note that we used an IDL-based nearest-neighbor linear interpolation routine to connect neighboring observations.  The evolution of each profile is clearly 
correlated with phase and is expected for a magnetically confined disk whose orientation with respect to us changes as the system rotates.   
Figure \ref{Be50halpha} illustrates that in the last two epochs of our observations, the H$\alpha$ line still exhibits a broad double-peaked profile, albeit the level of emission has decreased, especially on the blue-shifted of the profile.  The IR lines (Br-11 and Br-$\gamma$) similarly weaken during these last two epochs, especially the blue-shifted region of the H I line profiles.  Moreover, the red-shifted emission component seems to signifcantly broaden in the final two epochs, essentially merging with the residual lower intensity remnant of the blue-shifted profile.  Such behavior is not simply limited to these two IR lines, but rather occurs across all H I emission lines in our H-band data.

\subsection{Spectroscopy: HD 23478}\label{Be75specdiscussion}

\citet{sik15} recently presented a detailed optical spectropolarimetric study of HD 23478 in which they were able to both constrain the magnetic field 
properties of the system and confirm the presence of a magnetospherically confined disk via analysis of select optical absorption and emission lines.  
Our new 16 epochs of observations of the H I Br-11 line in the H-band help to fill in the phase-space explored by \citet{sik15}.  After continuum normalizing the spectra 
using standard techniques in IRAF, we computed EWs for our HD 23478 data, and phased them using the known 1.0498 day 
period and ephemeris \citep{jer93,sik15}.  Our measurements, compiled in Table \ref{tab:observation_table} and plotted in Figure \ref{Be75apogee}, exhibit clear phase-locked periodic behavior similar in morphology to that 
observed at H$\alpha$ (see e.g. Figure 11 of \citealt{sik15}).  

A differential dynamical spectrum for HD 23478, computed using our full 16 epochs of H I Br-11 data spanning $\sim$1.5 years and phased to the system's 1.0498 day period, illustrates that the overall 
morphology, especially the red-shifted side of the emission line profile, remained similar for more than 500 orbital periods.  There is some evidence for subtle changes in the line profile morphology 
between orbital periods via some of the ``choppiness'' in regions of the differential dynamical spectra with closely (in phase) spaced observations.  This might be indicative of small changes in the density of confined circumstellar gas over time.

\section{Discussion}
\label{sec:disc}

HD 345439 has been suspected to be a rapidly rotating $\sigma$ Ori E-type analog, based on the basic properties of its discovery SDSS III/APOGEE 
H-band spectrum and single-epoch follow-up spectra in the optical and IR \citep{eik14} as well as the suggestive presence of a rapidly variable 
magnetic field in the system \citep{hub15}.  Our analysis of $>$2500 epochs of photometry of the system from three independent surveys has 
consistently revealed the presence of a 0.7701 day photometric periodic signal (Figure \ref{kelt_fig}).   We note that the morphology of the light curve 
clearly deviates from a simple sinusoidal shape and instead exhibits evidence of twice per period dimmings of the central star, which is expected when 
magnetospherically confined disk material occults the central star.  Moreover, the observed deviation from simple sinusoidal phase symmetry has been observed 
in other $\sigma$ Ori E-type systems (e.g. \citealt{gru12,riv13,tow13}).  We can not however rule out that pulsations also play a role in producing the observed 
light curve morphology, particularly in light of the highly varying spectropolarimetric data of \citet{hub15}.  If however this 0.7701 day periodicity corresponds to HD 345439's rotational period as
we suggest, this would indicate HD 345439 is rotating significantly faster than the benchmark for this class of stars, the 1.19 day period of 
$\sigma$ Ori E \citep{tow10,tow13}.  Moreover, it would place HD 345439 instead closer to the fastest analogs of the class, HR 7355 
(0.52 days; \citealt{riv08,mik10,oks10}) and HR 5907 (0.508 days; \citealt{gru12}).  Our multi-epoch optical and IR spectra have confirmed that both the H I 
emission line strengths and profiles and He I absorption line strengths exhibit phase-locked variations on this same periodic time-scale, 
providing confirmation that the system does have circumstellar material confined in a disk-like geometry, likely by the suspected magnetic field \citep{hub15}.  As such, these data strongly support the original classification \citep{eik14} of HD 345439 as a $\sigma$ Ori E analog.  

Although detailed modeling is beyond the scope of this manuscript, we consider the broad properties of the HD 345439 system based on the 
observed light curve morphology and line profile behavior.  In particular, we qualitatively compare the properties of the HD 345439 system against 
the Rigidly Rotating Magnetosphere model \citep{to05b} and the resultant light curve \citep{tow08} and spectroscopic behavior predicted by this model.  
Using the interactive visualization movies by RRM model originator R. Townsend \footnote[12]{http://www.astro.wisc.edu/~townsend/}, we suggest the 
observed ``w-shaped'' light curve morphology (Figure \ref{kelt_fig}) and H I line profile evolution (Figure \ref{Be50halpha}) of HD 345439 is broadly consistent with the system having a moderately high inclination ($\sim$75$^{\circ}$) and a moderate 
obliquity ($\beta$ $\sim$45$^{\circ}$).  Detailed modeling of the system will require stronger constraints on the magnetic field properties of the system, 
beyond the observationally-based hint of a rapidly varying magnetic field in the system \citep{hub15}, as well as a better understanding of any anisotropies in its 
elemental surface abundances and the level to which they contribute to the observed broad-band light curves and He I EW behavior.

\citet{sik15} used spectropolarimetric observations to derive the longitudinal magnetic field strength and begin to parameterize the variable optical H I 
emission and He I absorption associated with the HD 23478 system.  Our IR spectroscopic data confirmed the presence of confined circumstellar gas by the 
system's magnetic field and provided enhanced phase coverage of diagnostics of this gas.  More broadly, our multiwavelength data advanced the exploration 
of optical and IR parameter space of $\sigma$ Ori E-type stars by \citet{oks15}.  As these multiwavelength data probe different optical depths in the disk, small-scale 
differences between near-contemporaneous optical and IR line profiles (e.g. the final two epochs of our optical and IR spectra of HD 345439; Figure \ref{Be50halpha}) could 
be used to help resolve the current uncertainties regarding the detailed morphology of magnetospherically confined disk material \citep{car13,ok15b}.

\vspace{1cm}

We thank our referee for providing us with helpful feedback that improved our paper.  We acknowledge funding from NSF-AST 1412110 (JPW).  NDR acknowledges his CRAQ (Qu\'ebec) postdoctoral fellowship.  
This paper makes use of data from the first public release of the WASP data (Butters et al. 2010) as provided by the WASP consortium and services at the NASA Exoplanet Archive, which is operated by the California Institute of Technology, under contract with the National Aeronautics and Space Administration under the Exoplanet Exploration Program.

\clearpage
}

\begin{sidewaystable}
\begin{tiny}
\vspace{-0.15in}
\center{\begin{tabular}{|l|c|c|c|c|c|c|c|c|c|c|c|}
\tableline
Name & Date & JD & Expos & H$\alpha$ & He I 4144 & He I 4387 & He I 4920 & He I 5015 & He I 6678 & Br-11 & phase \\
\tableline
\nodata & \nodata & \nodata & sec. & \AA\ & \AA\ & \AA\ & \AA\ & \AA\ & \AA\ & \AA\ & \nodata \\
\tableline
HD 345439 & 2012 Sept 2  & 2456172.8283 & 900 & -1.59$\pm$0.04  & 	1.91$\pm$0.03 &  1.87$\pm$0.04 & 1.42$\pm$0.02 &  0.58$\pm$ 0.01 & 1.20$\pm$0.02 & \nodata & 0.1111 \\
HD 345439 & 2012 Oct 4 & 2456204.8024 & 583 & 1.92$\pm$0.05 &  1.86$\pm$0.13 &  1.54$\pm$0.07  & 1.48$\pm$0.04 &  0.62$\pm$0.02 &  1.26$\pm$0.04 & \nodata & 0.7316 \\
HD 345439 & 2013 July 14 & 2456487.6805 & 900 & -1.09$\pm$0.03 & 1.56$\pm$0.04 &  1.52$\pm$0.03 & 1.41$\pm$0.02 &  0.58$\pm$0.01 & 1.16$\pm$0.03 &  \nodata & 0.9571 \\
HD 345439 & 2013 Jul 14 & 2456487.7243 & 1080 &  \nodata & \nodata & \nodata & \nodata & \nodata & \nodata & -4.1$^{T}$ $\pm$0.21 & 0.0139 \\
HD 345439 & 2013 July 14  & 2456487.7564 & 900 & -1.83$\pm$0.03 &  1.70$\pm$0.03 & 1.52$\pm$0.02 &  1.46$\pm$0.02 & 0.58$\pm$0.01 & 1.20$\pm$0.03 &  \nodata & 0.0556 \\
HD 345439 & 2013 Jul 14 & 2456487.7757 & 720 &  \nodata & \nodata & \nodata & \nodata & \nodata & \nodata & -5.7$^{T}$ $\pm$0.33 & 0.0807 \\
HD 345439 & 2013 July 14  & 2456487.8088 & 900 & -2.21$\pm$0.04 &  1.83$\pm$0.03 & 1.64$\pm$0.02 &1.45$\pm$0.02 &  0.57$\pm$0.01 & 1.24$\pm$0.02 & \nodata & 0.1237 \\
HD 345439 & 2013 Jul 14  & 2456487.8215 & 720 &  \nodata & \nodata & \nodata & \nodata & \nodata & \nodata & -5.7$^{T}$ $\pm$0.33 & 0.1401 \\
HD 345439 & 2013 July 14  & 2456487.8515 & 900 & -1.81$\pm$0.04 & 1.75$\pm$0.03 & 1.67$\pm$0.02 & 1.42$\pm$0.02 & 0.60$\pm$0.01 & 1.10$\pm$0.03 & \nodata & 0.1791  \\
HD 345439 & 2013 Jul 14  & 2456487.8681 & 720 &  \nodata & \nodata & \nodata & \nodata & \nodata & \nodata & -5.4$^{T}$ $\pm$0.31 & 0.2007 \\
HD 345439 & 2013 July 14  & 2456487.9007 & 900 & -1.16$\pm$0.02 & 1.64$\pm$0.03 &  1.47$\pm$0.03 & 1.25$\pm$0.02 & 0.53$\pm$0.01 & 1.06$\pm$0.02 & \nodata & 0.243 \\
HD 345439 & 2013 Jul 14  & 2456487.9153 & 720 &  \nodata & \nodata & \nodata & \nodata & \nodata & \nodata & -5.3$^{T}$ $\pm$0.34 & 0.2620 \\
HD 345439 & 2013 July 14  & 2456487.9525 & 900 & -0.40$\pm$0.02 & 1.78$\pm$ 0.04 & 1.44$\pm$0.04 & 1.30$\pm$0.02 & 0.51$\pm$0.01 & 1.03$\pm$0.03 & \nodata & 0.3103 \\
HD 345439 & 2013 Jul 14  & 2456487.9681 & 720 &  \nodata & \nodata & \nodata & \nodata & \nodata & \nodata & -2.3$^{T}$ $\pm$0.18 & 0.3305 \\
HD 345439 & 2013 Sept 22  & 2456557.7046 &1200 &  -0.78$\pm$0.02 & 1.48$\pm$0.02  & 1.36$\pm$0.03  & 1.40$\pm$0.02 & 0.47$\pm$0.01 & 1.07$\pm$0.02 &  \nodata & 0.8856 \\
HD 345439 & 2014 Apr 11  & 2456758.9467 & 1800 & -2.18$\pm$0.03 &  1.84$\pm$0.03 & 1.55$\pm$0.02 &  1.43$\pm$0.02 & 0.57$\pm$0.01 & 1.34$\pm$0.03 & \nodata &  0.2051 \\
HD 345439 & 2014 Apr 11  & 2456758.9738 & 1800 & -1.93$\pm$0.04 &  1.89$\pm$0.02 & 1.50$\pm$0.02 &  1.39$\pm$0.02 & 0.54$\pm$0.01 & 1.22$\pm$0.02 & \nodata & 0.2403 \\
HD 345439 & 2014 June 13  &  2456821.8155 & 1200 & 1.90$\pm$0.04 &  1.65$\pm$0.05 & 1.42$\pm$0.02 & 1.29$\pm$0.03 &  0.50$\pm$0.02 & 1.07$\pm$0.05 &  \nodata & 0.8423 \\
HD 345439 & 2014 June 13  &  2456821.8377 & 1800 & 1.31$\pm$0.03 &  1.54$\pm$0.03 & 1.29$\pm$0.04  & 1.19$\pm$0.02 & 0.49$\pm$0.01 & 0.89$\pm$0.03 &  \nodata & 0.8711 \\
HD 345439 & 2014 June 13  &  2456821.8641 & 1800 & -0.03$\pm$0.02 & 1.48$\pm$0.03 & 1.22$\pm$0.03  & 1.23$\pm$0.02 & 0.41$\pm$0.01 & 0.94$\pm$0.02 &  \nodata & 0.9054 \\
HD 345439 & 2014 June 13 &  2456821.8891 & 1800 & 0.03$\pm$0.02 &  1.32$\pm$0.03 &  1.32$\pm$0.02 & 1.24$\pm$0.02 & 0.44$\pm$0.01 & 0.91$\pm$0.02 & \nodata &  0.9378 \\
HD 345439 & 2014 June 13 &  2456821.9183 & 1800 & -0.96$\pm$0.03 & 1.51$\pm$0.02 & 1.28$\pm$0.02 & 1.36$\pm$0.02 & 0.55$\pm$0.01 & 1.00$\pm$0.03 & \nodata &  0.9758 \\
HD 345439 & 2014 June 13  &  2456821.9426 & 1800 & -1.08$\pm$0.03 &  1.53$\pm$0.03 & 1.46$\pm$0.02 & 1.28$\pm$0.02 &  0.49$\pm$0.01 & 1.19$\pm$0.03 & \nodata &  0.0073 \\
HD 23478 & 2012 Aug 29 & 2456168.9409	 &  2000 & \nodata & \nodata & \nodata & \nodata & \nodata & \nodata & -1.5$^{A}$ $\pm$0.02 & 0.055 \\
HD 23478 & 2012 Aug 31  & 2456170.8995	& 2000 &  \nodata & \nodata & \nodata & \nodata & \nodata & \nodata & -1.6$^{A}$ $\pm$0.02 & 0.921 \\
HD 23478 & 2012 Sep 1  & 2456171.9757	 &	 2500 &  \nodata & \nodata & \nodata & \nodata & \nodata & \nodata & -1.4$^{A}$ $\pm$0.02 & 0.946 \\
HD 23478 & 2012 Sep 2  & 2456172.9738 &	2000 &  \nodata & \nodata & \nodata & \nodata & \nodata & \nodata & -1.5$^{A}$ $\pm$0.02 &	0.897 \\
HD 23478 & 2012 Sep 4  & 2456174.9257 & 1000 &   \nodata & \nodata & \nodata & \nodata & \nodata & \nodata & -1.6$^{A}$  $\pm$0.03 & 	0.756 \\
HD 23478 & 2012 Nov 4 & 2456235.7739 &	2000 &  \nodata & \nodata & \nodata & \nodata & \nodata & \nodata & -1.8$^{A}$ $\pm$0.03 & 0.718 \\
HD 23478 & 2012 Dec 1  & 2456262.6834 & 2500 	&  \nodata & \nodata & \nodata & \nodata & \nodata & \nodata &  -1.1$^{A}$ $\pm$0.02 & 	0.351 \\
HD 23478 & 2012 Dec 21  &  2456282.6215	&	2000 &  \nodata & \nodata & \nodata & \nodata & \nodata & \nodata &  -0.6$^{A}$ $\pm$0.01 & 0.343 \\
HD 23478 & 2012 Dec 22 & 2456283.6242 & 	2000 &   \nodata & \nodata & \nodata & \nodata & \nodata & \nodata & -1.1$^{A}$ $\pm$0.02 & 	0.298 \\
HD 23478 &2013 Sep 26  & 2456561.9916 & 1500 &  \nodata & \nodata & \nodata & \nodata & \nodata & \nodata & -0.9$^{A}$ $\pm$0.02 &	0.460 \\
HD 23478 & 2014 Jan 12  & 2456669.6736 & 2000 &  \nodata & \nodata & \nodata & \nodata & \nodata & \nodata & -1.2$^{A}$ $\pm$0.02 & 	0.034 \\
HD 23478 & 2014 Jan 16  & 2456673.5636 & 2000 &  \nodata & \nodata & \nodata & \nodata & \nodata & \nodata & -0.9$^{A}$ $\pm$0.02 &	0.740 \\
HD 23478 & 2014 Jan 18  & 2456675.5648	&	2000 &  \nodata & \nodata & \nodata & \nodata & \nodata & \nodata & -1.1$^{A}$ $\pm$0.02 & 0.646 \\
HD 23478 & 2014 Feb 13  & 2456701.5886	&	2000 &  \nodata & \nodata & \nodata & \nodata & \nodata & \nodata & -0.8$^{A}$ $\pm$0.02 & 0.435 \\
HD 23478 & 2014 Feb 15  & 2456703.5901 &	2000 &  \nodata & \nodata & \nodata & \nodata & \nodata & \nodata & -0.8$^{A}$ $\pm$0.02 & 0.342 \\
HD 23478 & 2014 Feb 18  & 2456706.6017 & 	1000 &  \nodata & \nodata & \nodata & \nodata & \nodata & \nodata & -0.8$^{A}$ $\pm$0.02 & 0.210 \\
\tableline
\end{tabular}}
\vspace{0.1in}
\caption{Summary of Spectroscopic Data:  The summary of our spectroscopic observations of HD 345439 and HD 23478.  Note that the ``A'' designation in column 11 corresponds to data obtained with the APOGEE instrument on the APO 2.5m while the ``T'' designation in column 11 corresponds to data obtained with the Triplespec instrument on the APO 3.5m.  The 
Julian Dates (JDs) listed in column 3 were computed based on the mid-exposure times of each observation.  The phases compiled in column 12 were computed using the 0.7701 day period and ephemeris derived in Section \ref{Be50photdiscussion} for HD 345439 and using the previously reported 1.0498 day period and ephemeris for HD 23478 \citep{jer93,sik15}.  \label{tab:observation_table}}
\end{tiny}
\end{sidewaystable}

\begin{figure*}
\center{\includegraphics[width=1.0\textwidth]{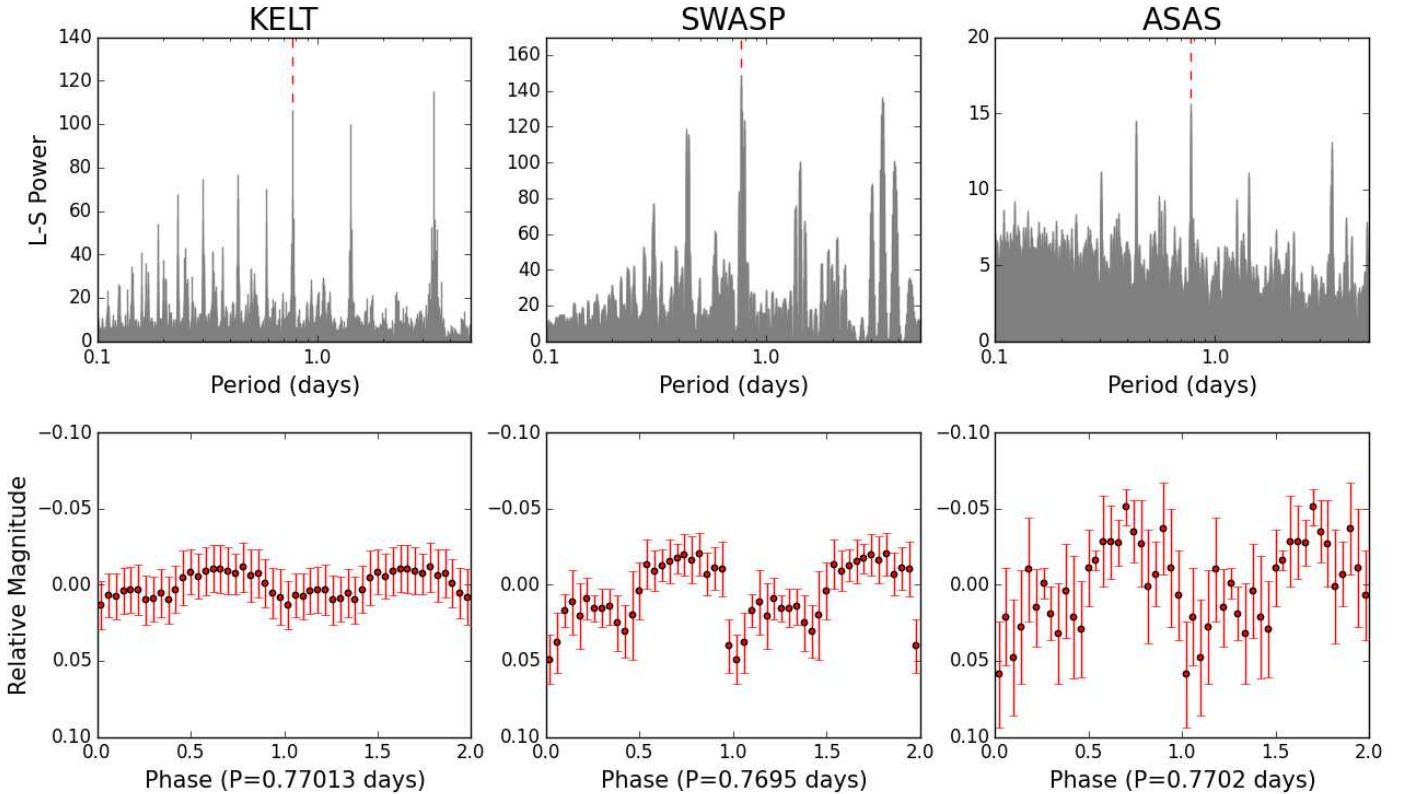}}
\caption{The available archival photometric data for HD 345439 all exhibit similar peaks in their Lomb-Scargle periodograms (KELT: 0.7701 days, upper left panel; SuperWASP: 0.7695 days, 
upper middle panel; ASAS: 0.7702 days, upper right panel).  Phase-folding by these periods and computing median flux values in bins of 0.05 phase reveals similar morphologies for all 
three data sets (KELT, lower left panel; SuperWASP, lower middle panel; ASAS, lower right panel).}
\label{kelt_fig}
\end{figure*}

\newpage
\begin{figure*}
\center{\includegraphics[width=0.7\textwidth]{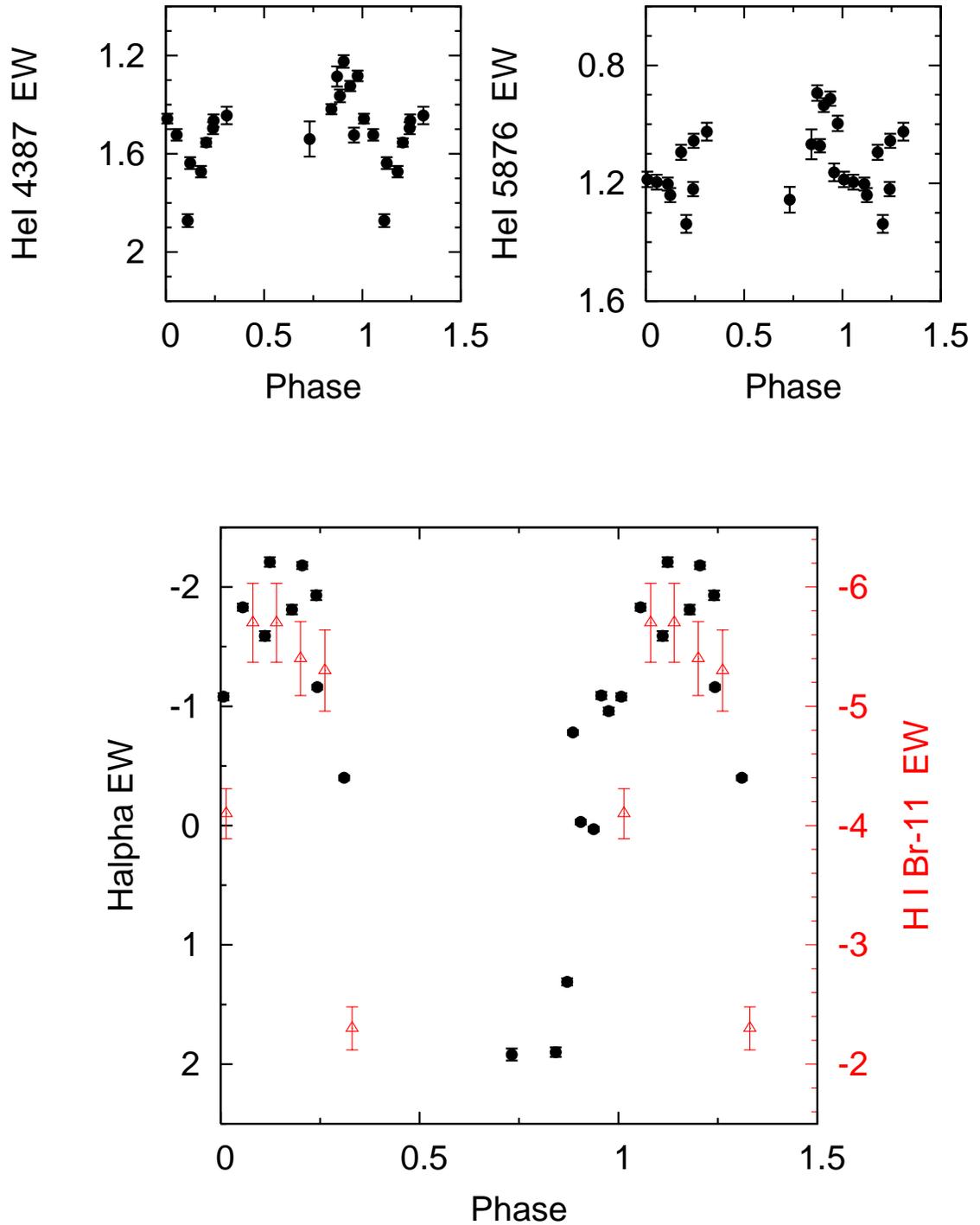}}
\caption{The equivalent widths of select lines in HD 345439, compiled in Table \ref{tab:observation_table}, are plotted as a function of phase.  These lines all exhibit evidence 
of phase-locked variations in line strengths, as do additional He I lines compiled in Table \ref{tab:observation_table} but not plotted here (He I 4144 \AA , He I 4920 \AA , 
He I 5015 \AA , and He I 6678 \AA ).  While our phase coverage of the H I Br-11 line (red triangles, bottom panel) is more minimal compared to other lines, these 
data still mimic the general behavior observed in the H$\alpha$ line (black circles, bottom panel).}
\label{Be50phase}
\end{figure*}

\clearpage
\newpage
\begin{figure*}
\center{\includegraphics[width=0.3\textwidth]{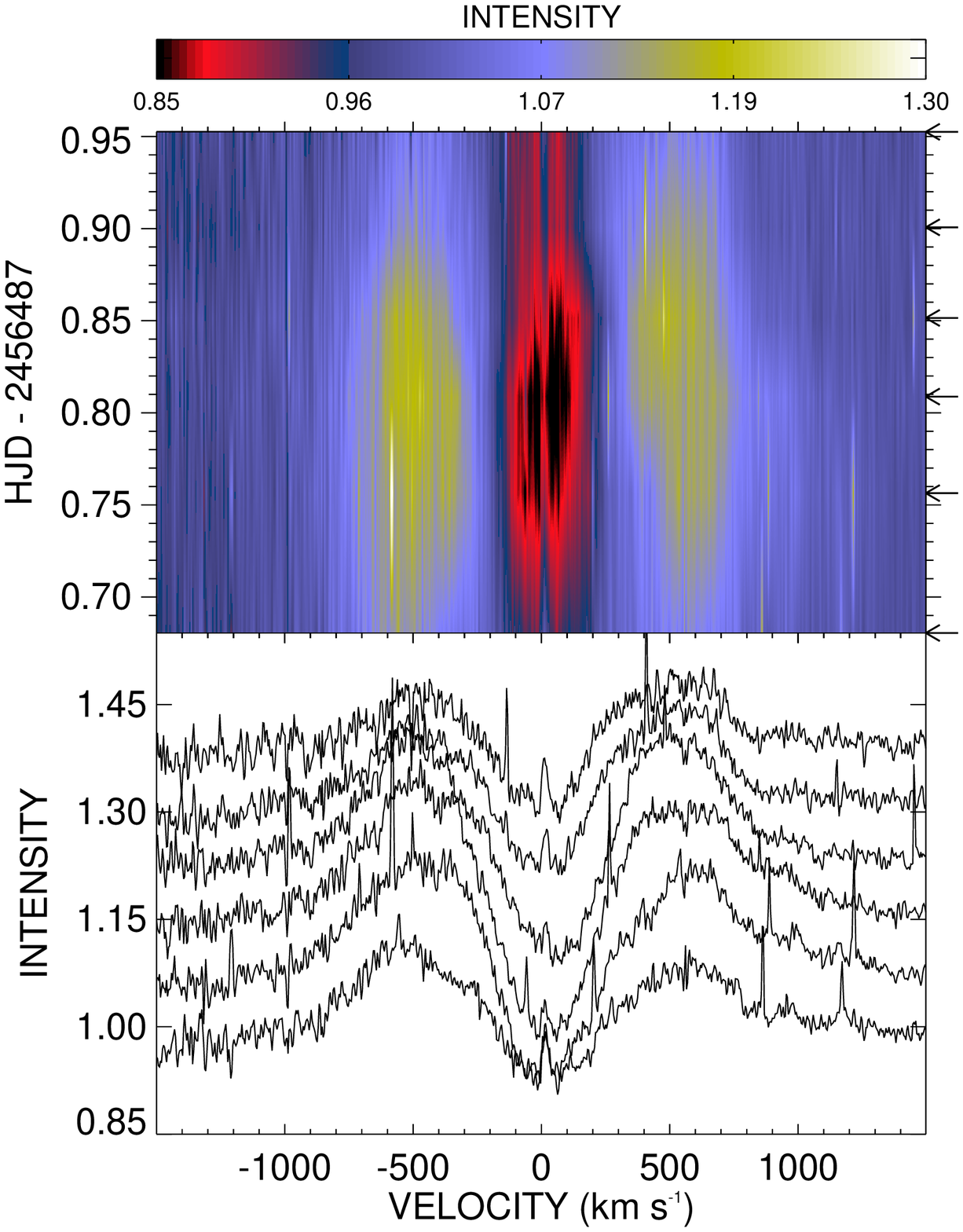}\includegraphics[width=0.3\textwidth]{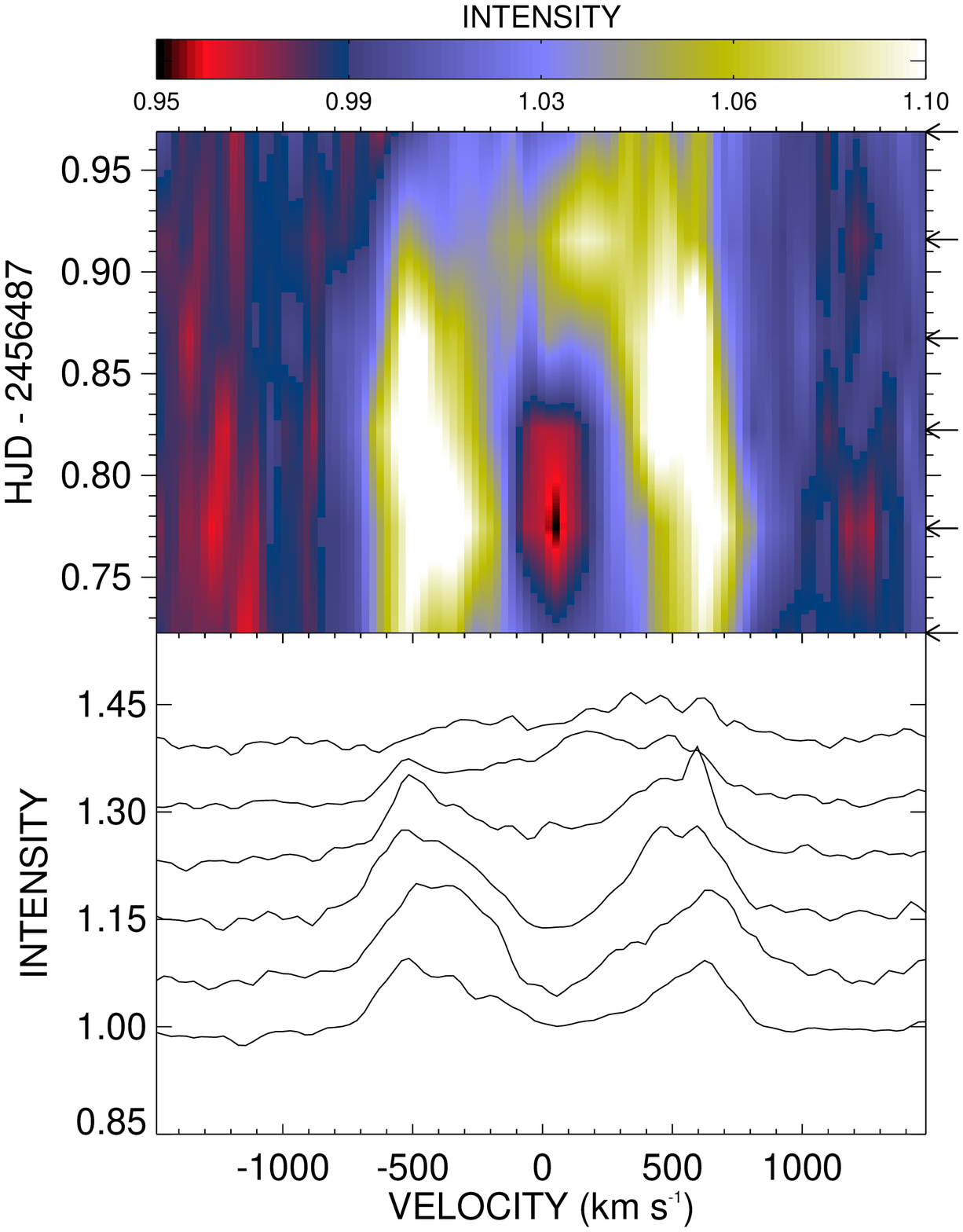}\includegraphics[width=0.3\textwidth]{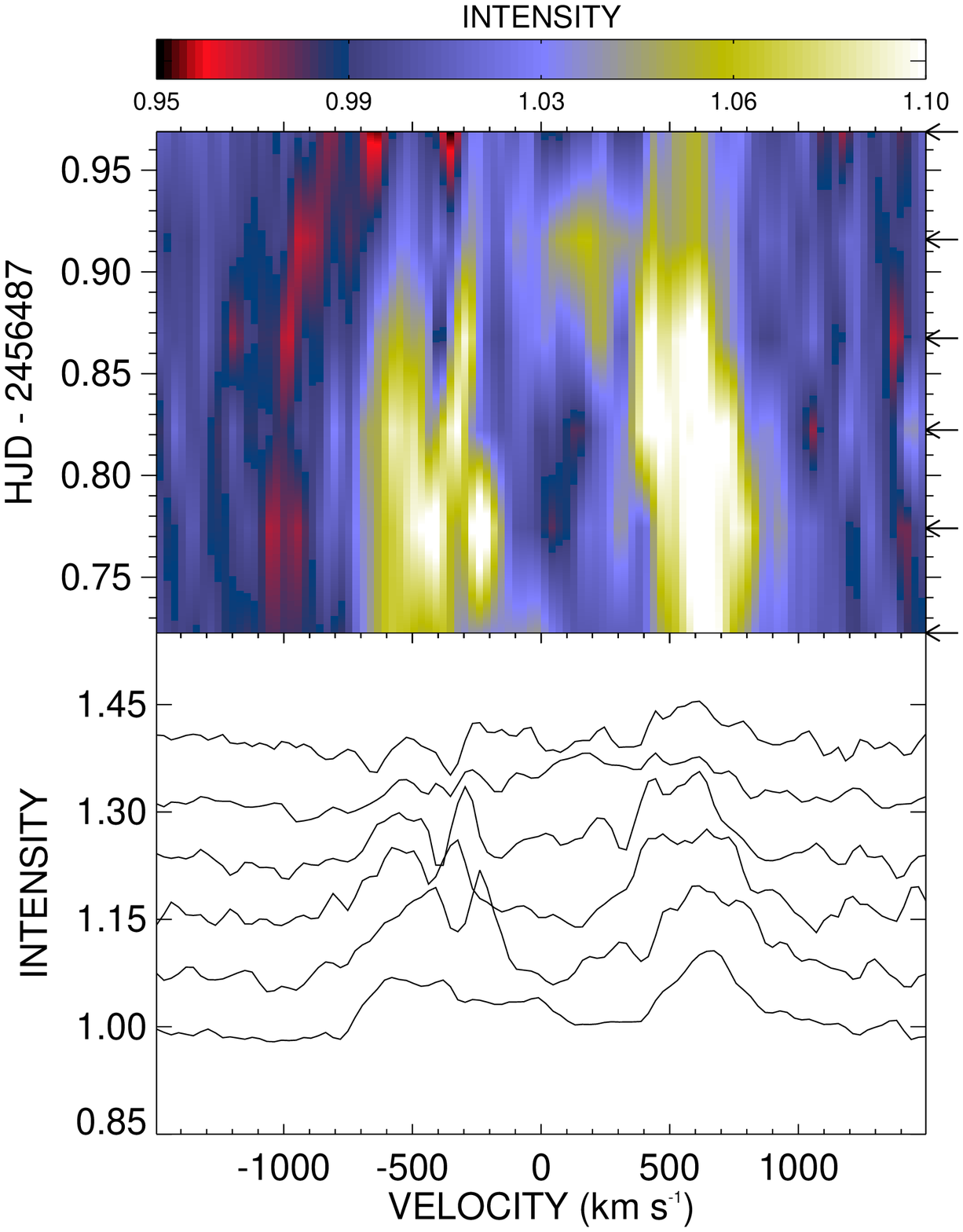}}
\caption{Dynamical spectra of the HD 345439 H$\alpha$ (left panels), H I-Br11 (center panels), and H I-Br$\gamma$ (right panels) lines from the six epochs of optical and six epochs of IR 
spectra obtained on 2013 July 14 show the clear evolution of these emission line profiles over the coarse of $\sim$40\% of the system's rotational 
period (from 0.95 - 0.33).  The arrows on the right-side of each top panel depicts the epoch of each exposure, while the bottom panel depicts a zoomed view of each line profile.}
\label{Be50halpha}
\end{figure*}

\clearpage
\newpage
\begin{figure*}
\center{\includegraphics[width=0.4\textwidth]{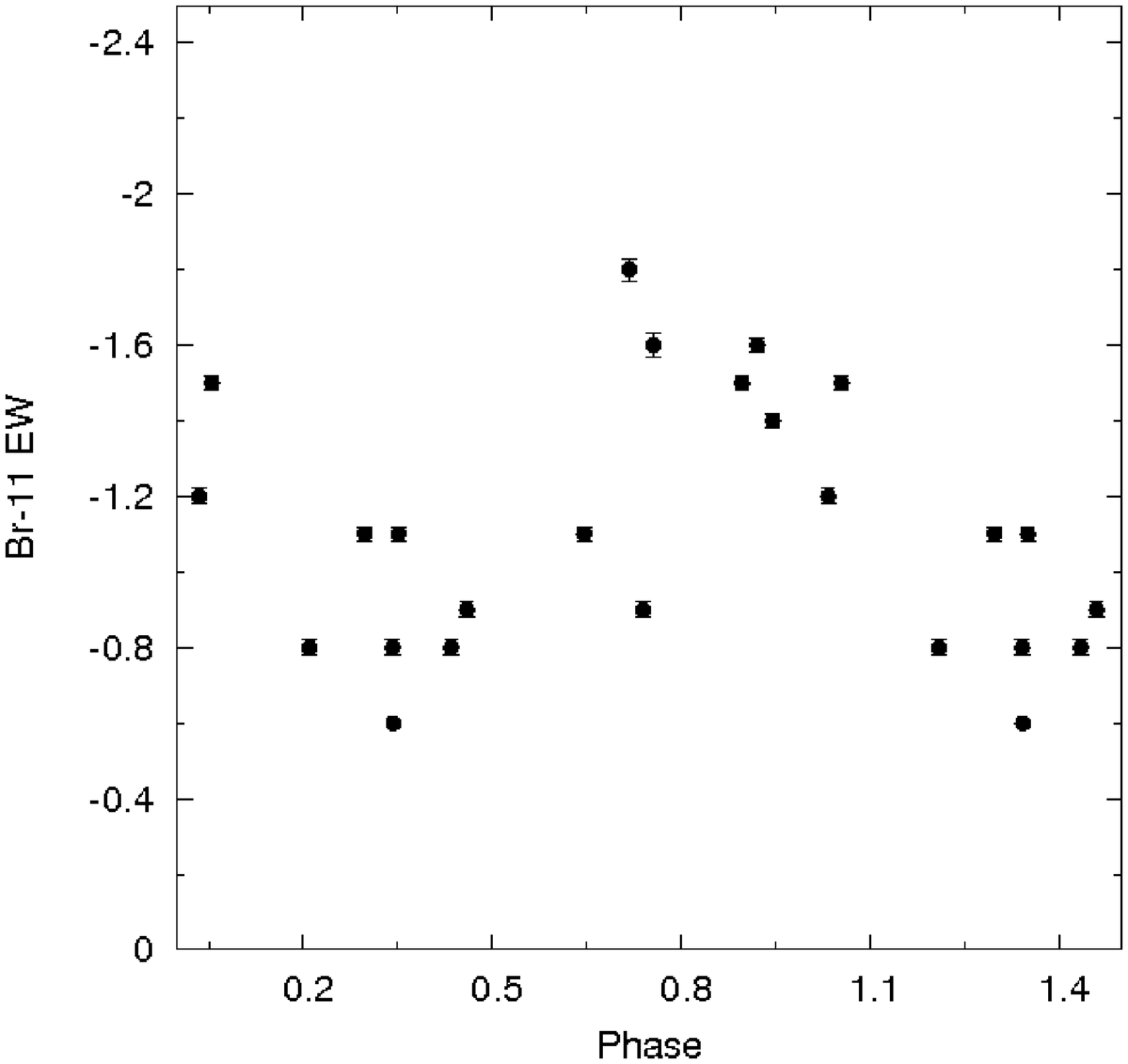}\includegraphics[width=0.4\textwidth]{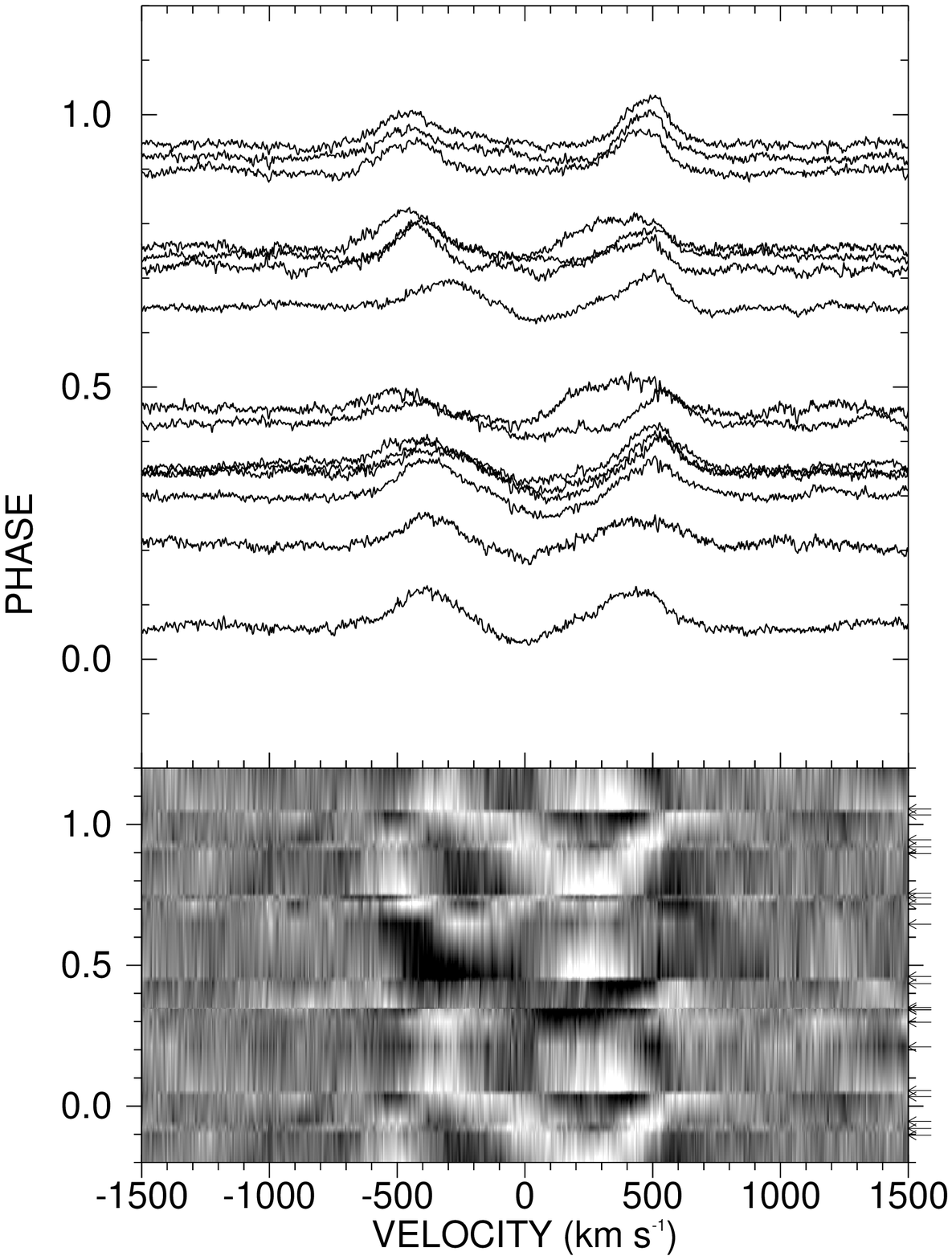}}
\caption{The EWs of H I Br-11 our HD 23478 data exhibit a clear phase-locked periodic behavior at the known 1.0498 
period of the system \citep{jer93,sik15}, and 
exhibit a similar morphology to that reported at H$\alpha$ by \citet{sik15}.  The differential dynamical spectra of HD 23478 also exhibit a phase-locked 
variability pattern, which is especially visible in the red-shifted side of the emission line profile over phase. Note that 
the intensity scaling of these differential dynamical spectra is $\pm$2.5\%.}
\label{Be75apogee}
\end{figure*}

\end{document}